\documentclass[fleqn,10pt]{wlscirep}
\usepackage[utf8]{inputenc}
\usepackage[T1]{fontenc}
\usepackage{amsmath}
\usepackage{multirow}
\usepackage{subcaption}
\usepackage{graphicx}
\usepackage{hyperref}
\usepackage{url}
\usepackage{xcolor}
\usepackage{multicol}
\usepackage{float}
\usepackage{caption}
\usepackage{algorithmic,algorithm}
\usepackage[sectionbib]{chapterbib}
\usepackage{verbatim}
\usepackage{stfloats}
\usepackage{cite}
\usepackage[super, comma, compress]{natbib}
\begin{document}

\title{DRAC: Diabetic Retinopathy Analysis Challenge with Ultra-Wide Optical Coherence Tomography Angiography Images}

% \cortext[cor1]{Corresponding authors.}
%\fnref{fn1}
\author[1]{Bo Qian}
\author[2,3]{Hao Chen}
\author[4]{Xiangning Wang}
\author[2]{Haoxuan Che}
\author[5]{Gitaek Kwon}
\author[5]{Jaeyoung Kim}
\author[6]{Sungjin Choi}
\author[6]{Seoyoung Shin}
\author[7]{Felix Krause}
\author[7]{Markus Unterdechler}
\author[8]{Junlin Hou}
\author[8,9]{Rui Feng}
\author[10,11]{Yihao Li}
\author[10,11]{Mostafa El Habib Daho}
\author[4]{Qiang Wu}
\author[12,13,14]{Ping Zhang}
\author[15]{Xiaokang Yang}
\author[16]{Yiyu Cai}
\author[17,18]{Huating Li}
\author[17,18]{Weiping Jia}
\author[1,15,*]{Bin Sheng}
% \cortext[cor1]{Corresponding authors.}
% \ead{shengbin@sjtu.edu.cn}

\affil[1]{Department of Computer Science and Engineering, Shanghai Jiao Tong University, Shanghai, China}
\affil[2]{Department of Computer Science and Engineering, The Hong Kong University of Science and Technology, Hong Kong, China}
\affil[3]{Department of Chemical and Biological Engineering, The Hong Kong University of Science and Technology, Hong Kong, China.}
\affil[4]{Department of Ophthalmology, Shanghai Jiao Tong University Affiliated Sixth People’s Hospital, Shanghai, China}
\affil[5]{VUNO Inc., Seoul, Korea}
\affil[6]{AI/DX Convergence Business Group, KT, Korea}
\affil[7]{Johannes Kepler University Linz, Linz, Austria}
\affil[8]{School of Computer Science, Shanghai Key Laboratory of Intelligent Information Processing, Fudan University, China}
\affil[9]{Academy for Engineering and Technology, Fudan University, China}
\affil[10]{LaTIM UMR 1101, Inserm, Brest, France}
\affil[11]{University of Western Brittany, Brest, France}
\affil[12]{Department of Computer Science and Engineering, The Ohio State University, Ohio, USA}
\affil[13]{Department of Biomedical Informatics, The Ohio State University, Ohio, USA}
\affil[14]{Translational Data Analytics Institute, The Ohio State University, Ohio, USA}
\affil[15]{MoE Key Lab of Artificial Intelligence, Artificial Intelligence Institute, Shanghai Jiao Tong University, Shanghai, China}
\affil[16]{School of Mechanical and Aerospace Engineering, Nanyang Technological University, Singapore}
\affil[17]{Department of Endocrinology and Metabolism, Shanghai Jiao Tong University Affiliated Sixth People’s Hospital, Shanghai, China}
\affil[18]{Shanghai Diabetes Institute, Shanghai Clinical Center for Diabetes, Shanghai, China}
\affil[*]{Correspondence: shengbin@sjtu.edu.cn (B.S.)}

\begin{abstract}
Computer-assisted automatic analysis of diabetic retinopathy (DR) is of great importance in reducing the risks of vision loss and even blindness. Ultra-wide optical coherence tomography angiography (UW-OCTA) is a non-invasive and safe imaging modality in DR diagnosis system, but there is a lack of publicly available benchmarks for model development and evaluation. To promote further research and scientific benchmarking for diabetic retinopathy analysis using UW-OCTA images, we organized a challenge named "DRAC - Diabetic Retinopathy Analysis Challenge" in conjunction with the 25th International Conference on Medical Image Computing and Computer Assisted Intervention (MICCAI 2022). The challenge consists of three tasks: segmentation of DR lesions, image quality assessment and DR grading. The scientific community responded positively to the challenge, with 11, 12, and 13 teams from geographically diverse institutes submitting different solutions in these three tasks, respectively. This paper presents a summary and analysis of the top-performing solutions and results for each task of the challenge. The obtained results from top algorithms indicate the importance of data augmentation, model architecture and ensemble of networks in improving the performance of deep learning models. These findings have the potential to enable new developments in diabetic retinopathy analysis. The challenge remains open for post-challenge registrations and submissions for benchmarking future methodology developments.
\end{abstract}
%\begin{document}
\twocolumn
\flushbottom
\maketitle
\thispagestyle{empty}

\section{Introduction}
Diabetic retinopathy (DR) is one of the most common complications caused by diabetes \citep{reichel2015diabetic}. Patients with DR are more likely to get vision impairment and even blindness than healthy individuals. DR affects a large amount of the working-age population worldwide. According to the International Diabetes Federation \citep{atlas2017brussels}, it is estimated that about 700 million people in the world are expected to have diabetes by 2045, and one-third of them will have diabetic retinopathy. DR is diagnosed by visually inspecting retinal fundus images for the presence of retinal lesions, such as microaneurysms (MAs), intraretinal microvascular abnormalities (IRMAs) and neovascularization (NV) \citep{wang2018diabetic}. Hence, the detection of these lesions is significant for DR diagnosis.

Regular DR screening and timely treatment can be implemented to reduce the risks of vision loss and blindness \citep{american202011}. However, the whole population screening faces many challenges. First, comprehensive DR screening puts a heavy burden on ophthalmologists. Especially in developing countries and rural parts, there may not be enough medical resources and ophthalmologists to perform the DR screening \citep{jones2010diabetic, lin2016addressing, wang2017availability, ting2016diabetic}. Second, DR screening relies heavily on the experience of ophthalmologists. Differences in the experience of professional ophthalmologists may lead to different diagnoses, and the inadequate training of ophthalmologists can also result in misdiagnosis and low accuracy in DR screening \citep{dai2021deep, wang2017availability}. Third, systematic DR screening is associated with complicated social management and economic burden. Therefore, an effective computer-aided system is essential to assist manual DR screening, further providing accurate diagnosis and reasonable treatment, thus reducing the burden on ophthalmologists \citep{jelinek2009automated, li2018automated, gulshan2016development}. The commonly used imaging modalities in DR diagnosis include fundus photography, fluorescein angiography (FA) and optical coherence tomography angiography (OCTA). Fundus photography is useful in recording the distribution of hard exudates, and retinal alterations in severe non-proliferative diabetic retinopathy (NPDR), but is difficult to detect early or small neovascular lesions. FA is mainly used to detect the presence of neovascularization, but it is an invasive fundus imaging and cannot be used in patients with allergies, pregnancy, and poor kidney function. OCTA can non-invasively identify the changes of DR neovascularization and help ophthalmologists diagnose proliferative diabetic retinopathy (PDR). Further, ultra-wide OCTA (UW-OCTA) can show a wider area around the retina that is not captured by typical OCTA, making it an important imaging modality in the diagnosis of DR.

The grand challenge plays an important role in promoting the application of deep learning methods in medical image analysis. It defines one or more clinically relevant tasks and provides the corresponding dataset, encouraging the participating teams to develop state-of-the-art algorithms for the tasks and make a fair comparison. Many challenges have been organized for the analysis of diabetic retinopathy, such as ROC \citep{niemeijer2009retinopathy}, IDRiD \citep{porwal2020idrid} and DeepDRiD \citep{liu2022deepdrid}. The imaging modality used in these challenges is fundus photography. As far as we know, currently no challenge has been attempted to explore UW-OCTA imaging modality for DR analysis. To address above limitations, we organized the Diabetic Retinopathy Analysis Challenge (DRAC) at MICCAI 2022, which allowed for the first time to make an analysis of method performance for diabetic retinopathy using UW-OCTA imaging modality. Three clinically relevant tasks are proposed in DRAC challenge, including the segmentation of DR lesions, image quality assessment and DR grading. The dataset and manual annotations for training are publicly available as an ongoing benchmarking resource at DRAC challenge website drac22.grand-challenge.org.

In this paper, we describe in detail the challenge setup, the released dataset, and the competing solutions that performed well on the challenge tasks. These solutions were partly presented at the satellite event of MICCAI 2022. We hope the algorithms described in the challenge will be of value to the diabetic retinopathy research community. The rest of the paper is organized as follows. We describe the challenge setup, dataset and evaluation in Section 2. We present a summary of the competing solutions from the participating teams in Section 3. We report and discuss the challenge results in Section 4. We discuss the characteristics of the competing solutions, significance of UW-OCTA images for DR analysis, limitations of the study and future work in Section 5. Finally, the conclusion is presented in Section 6.

\begin{figure}[!ht]
	\begin{center}
		\includegraphics[width=3.4in]{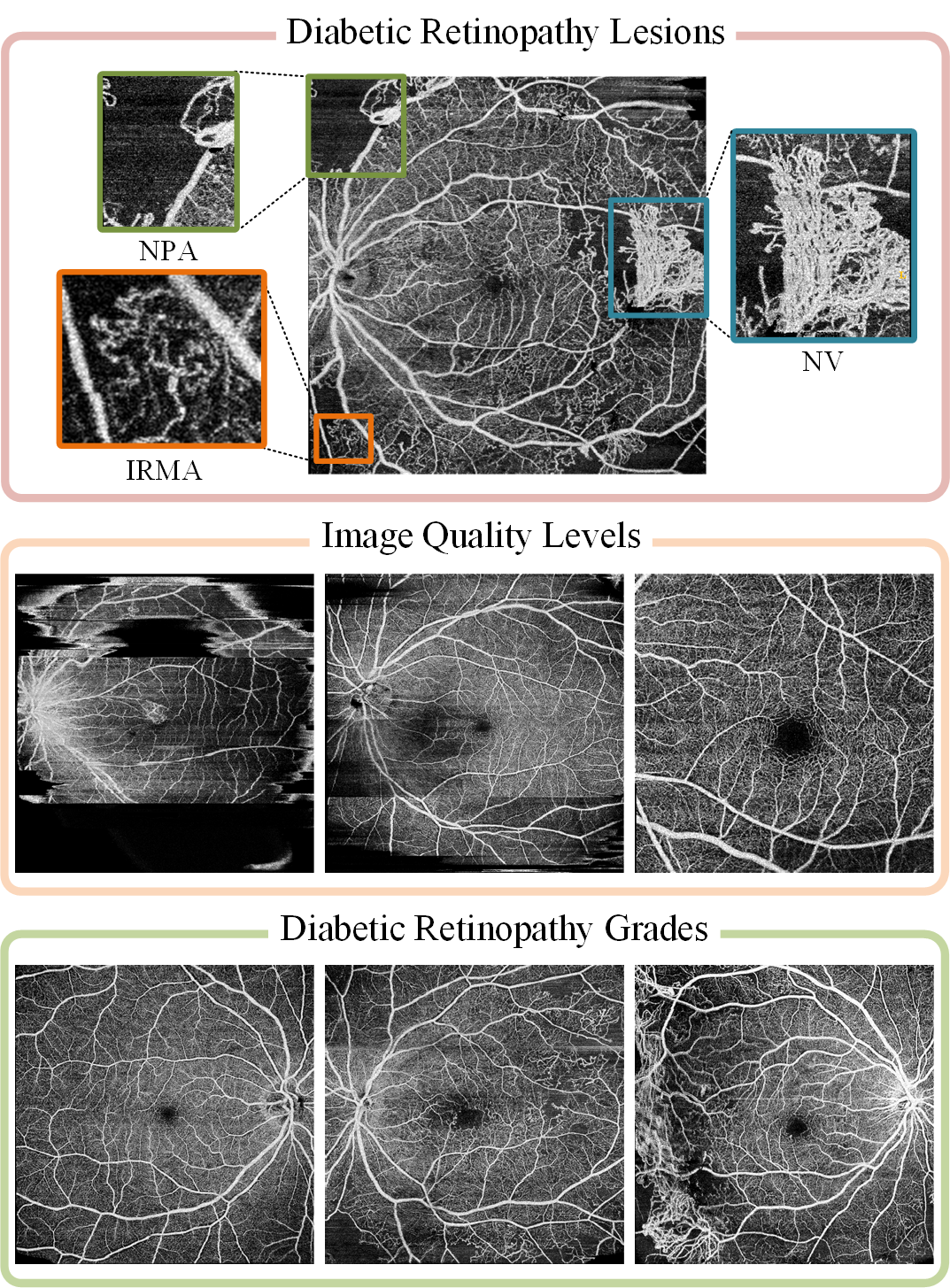}
		\caption{Examples of UW-OCTA images in three tasks of the challenge. The first row shows three different lesions in the task of DR lesion segmentation. The second row shows the different image quality levels in the task of image quality assessment. From left to right are the images with poor quality level, good quality level and excellent quality level respectively. The third row shows the different DR grades in the task of DR grading. From left to right are the images with non-DR, NPDR and PDR respectively. Abbreviation: intraretinal microvascular abnormality (IRMA), nonperfusion area (NPA), neovascularization (NV).}
		\label{example}
	\end{center}
\end{figure}

\section{Challenge description}
\subsection{Challenge setup}
DRAC challenge aimed at providing a benchmark for evaluating the algorithms that are used for the automatic diabetic retinopathy analysis using UW-OCTA images. It addressed the current lack of publicly available UW-OCTA datasets for fair performance evaluation of DR diseases. The challenge was subdivided into three tasks as follows. Fig \ref{example} shows the example images in each task.
\begin{enumerate}
\item Task 1: segmentation of DR lesions. There are three types of lesions to be segmented, including intraretinal microvascular abnormality (IRMA), nonperfusion area (NPA) and neovascularization (NV).
\item Task 2: image quality assessment. Classification of the image quality levels that  poor quality level, good quality level and excellent quality level.
\item Task 3: DR grading. Classification of DR grades according to the severity level of diabetic retinopathy, including non-DR, non-proliferative diabetic retinopathy (NPDR) and proliferative diabetic retinopathy (PDR).
\end{enumerate}

The organization of the challenge refers to the BIAS (Biomedical Image Analysis ChallengeS) guideline \citep{maier2020bias}. The challenge was officially announced at MICCAI 2022 and was hosted on Grand Challenge. On the challenge website, the participants can have access to the dataset after they registered on the website and signed the challenge rule agreement. In addition, participants can browse the challenge rules and news, submit results and find their rankings on the challenge website. In the case of multiple submissions, only the most recent run was counted for the final challenge result. We also provided submission guidelines for the participants. The challenge was launched in July 2022 by releasing the training dataset. The test set was released on 8$th$ August 2022, and the challenge submission was opened between 8$th$ August 2022 and 12$th$ September 2022. Each participating team was required to submit a method description paper with a minimum of 4 pages before 8$th$ October 2022. When participating in multiple tasks, each team can either submit several papers or a single paper reporting all methods. Finally, a total of 91 teams and individuals from more than 25 different countries or regions signed the challenge rule agreement consent form and downloaded the dataset throughout the challenge. Out of them, 17 teams submitted a total of 18 method description papers before the deadline, where 11, 12 and 13 different methods were reported for the three tasks respectively. Some teams participated in two or more tasks and choose to report their methods in one single paper. During the satellite event at MICCAI 2022 on 18$th$ September 2022, we summarized the challenge results and invited the top-ranked teams to present their algorithms.

\subsection{Challenge dataset}
A total of 1103 UW-OCTA images are collected in this challenge. The images have a resolution of 1024 $\times$ 1024 pixels and are stored in gray $png$ format. In the process of image annotating, many factors affect the image quality \citep{kawai2021prevention, wang2021deep}, such as artifacts, vascular quality, etc. The specific annotation standard of image quality assessment is shown in Table \ref{quality}. For DR grading, the specific grading standard for non-DR, NPDR and PDR refers to the international clinical diabetic retinopathy (DR) severity scale \citep{world2006prevention}. The data division method for each task is as follows. For image quality assessment task, the images are split into 60\% for training (665 images) and 40\% for testing (438 images). For DR grading task and DR lesion segmentation task, the data division method was the same as the task of image quality assessment, but in DR grading task, we removed the images with the poor quality levels and images that caused considerable controversies about the DR grade by ophthalmologists, then the rest images are retained as the training set (611 images) and test set (386 images) respectively. In DR lesion segmentation task, we only retained images that show representative lesions to form the training set (109 images) and test set (65 images). The ground truth of training set for the classification task was stored and provided in a $csv$ file. For the segmentation task, the ground truth was provided in the form of binary masks stored in $png$ format.

\begin{table*}[!htb]
\centering
\caption{Detailed descriptions for each image quality level.}
\begin{tabular}{llll}
\hline
Image quality level & Overall quality                         & Nonvascular area quality & Vascular quality         \\ \hline
Poor                & Insufficient                            & Severe artifacts         & Blurring                 \\
Good                & Moderate blurring or stripe noise       & Moderate Artifacts       & Moderate blurring        \\
Excellent           & Slight blurring or with slight stripe noise & Slight artifacts   & Clear or slight blurring \\ \hline
\end{tabular}
\label{quality}
\end{table*}

\begin{table*}[!htb]
\centering
\caption{List and details of the top three teams in each task of the challenge. We use A to represent the algorithm in task 1 (DR lesion segmentation), B to represent the algorithm in task 2 (image quality assessment), and C to represent the algorithm in task 3 (DR grading). The followed number denotes the ranking. For example, A1 denotes the first ranked algorithm in task 1, B1 denotes the first ranked algorithm in task 2, and C1 denotes the first ranked algorithm in task 3.}
\begin{tabular}{cclll}
\hline
\multicolumn{1}{l}{}    & Algorithm & Team name          & Affiliation                                    & Score  \\ \hline
\multirow{3}{*}{Task 1} & A1        & FAI                & VUNO Inc., Seoul, South Korea                  & 0.6067 \\
                        & A2        & KT\_Bio\_Health    & AI/DX Convergence Business Group, KT, Korea    & 0.6046 \\
                        & A3        & AILS\_Students@JKU & Johannes Kepler University Linz, Linz, Austria & 0.5756 \\ \hline
\multirow{3}{*}{Task 2} & B1        & FAI                & VUNO Inc., Seoul, South Korea                  & 0.8090 \\
                        & B2        & KT\_Bio\_Health    & AI/DX Convergence Business Group, KT, Korea    & 0.8075 \\
                        & B3        & FDVTS\_DR          & Fudan University, Shanghai, China              & 0.7896 \\ \hline
\multirow{3}{*}{Task 3} & C1        & FAI                & VUNO Inc., Seoul, South Korea                  & 0.8910 \\
                        & C2        & KT\_Bio\_Health    & AI/DX Convergence Business Group, KT, Korea    & 0.8902 \\
                        & C3        & LaTIM              & LaTIM UMR 1101, Inserm, Brest, France          & 0.8761 \\ \hline
\end{tabular}
\label{top3}
\end{table*}

\subsection{Evaluation and ranking}
There are three leaderboards on website and each task corresponds to a leaderboard. In the task of DR lesion segmentation, Dice Similarity Coefficient (DSC) is used for the algorithm evaluation and ranking in segmentation task. In case of a tie, Intersection of Union (IoU) is used as an auxiliary ranking metric. The DSC and IoU are calculated as follows.
	\begin{equation}
			DSC=\frac{2|G \cap P|}{|G|+|P|}
	\end{equation}
	\begin{equation}
			IoU=\frac{|G \cap P|}{|G \cup P|}
	\end{equation}
where $G$ and $P$ are the ground truth mask and predicted mask respectively. The metric is calculated for each class independently and then results are then averaged. For the tasks of image quality assessment and DR grading, quadratic weighted kappa is used for the algorithm evaluation and ranking. In case of a tie, Area Under ROC Curve (AUC) is used as an auxiliary ranking metric. The quadratic weighted kappa $K_w$ is calculated as follows.
    \begin{equation}
        K_w = 1-\dfrac{\sum_{i,j}{\text{w}_{i,j} O_{i,j}}}{\sum_{i,j}{\text{w}_{i,j} E_{i,j}}}
    \end{equation}
where $O$ and $E$ are the histogram matrix and expected matrix respectively, with the size of $N \times N$. The weighted matrix $\text{w}$ is defined by $\text{w}_{i,j}=\frac{(i-j)^2}{(N-1)^2}$, where $i$ and $j$ denote actual value and predicted value respectively. $N$ is the number of classes.

\section{Summary of competing solutions}
In this section, we present a summary of the top-ranked algorithms in each task of the challenge from the perspective of data pre-processing, data augmentation, model architecture, model optimization and post-processing. Specifically, we only summarize the top three best-performing algorithms in each task to keep the paper concise. The detailed information of the top three teams in each task is shown in Table \ref{top3}. For the convenience of description, we use A, B and C to denote the algorithms in task 1 (DR lesion segmentation), task 2 (image quality assessment) and task 3 (DR grading), respectively, followed by a number indicating the ranking of the algorithm in this task. For example, A1, B1, and C1 represent the first ranked algorithms in task 1, task 2, and task 3, respectively.

\begin{table*}[!htb]
\centering
\caption{Summary of data augmentation techniques of top three algorithms in each task of the challenge. Abbreviation: Flipping (F), Rotation (R), Cropping (C), Scaling (S), Brightness (B), Contrast (CT), Gamma (G), Sharpen (SP), Blur (BL), Grid Distortion (GD), Coarse Dropout (CD), Cut Out (CO), Gaussian Noise (GN), Affine (A), MixUp (MU), CutMix (CM), Test-Time Augmentation (TTA).}
\begin{tabular}{ccccccccccccl}
\hline
\multicolumn{1}{l}{}    & \multirow{2}{*}{Algorithm} & \multicolumn{11}{c}{Data augmentation}                                                                                                    \\ \cline{3-13} 
\multicolumn{1}{l}{}    &                            & F          & R          & C          & S          & B          & CT         & G          & SP         & BL         & TTA        & Others  \\ \hline
\multirow{3}{*}{Task 1} & A1                         & \checkmark & \checkmark &            & \checkmark & \checkmark & \checkmark & \checkmark & \checkmark & \checkmark & \checkmark & A/GD/CD \\
                        & A2                         & \checkmark & \checkmark &            &            &            &            &            &            &            &            &         \\
                        & A3                         & \checkmark & \checkmark &            & \checkmark & \checkmark & \checkmark & \checkmark & \checkmark & \checkmark &            & CO/GN   \\ \hline
\multirow{3}{*}{Task 2} & B1                         & \checkmark & \checkmark &            & \checkmark & \checkmark & \checkmark & \checkmark & \checkmark & \checkmark & \checkmark &         \\
                        & B2                         & \checkmark &            &            &            &            &            &            &            &            &            &         \\
                        & B3                         & \checkmark &            & \checkmark & \checkmark & \checkmark & \checkmark &            &            &            &            & MU/CM   \\ \hline
\multirow{3}{*}{Task 3} & C1                         & \checkmark & \checkmark &            & \checkmark & \checkmark & \checkmark & \checkmark & \checkmark & \checkmark & \checkmark & CD      \\
                        & C2                         & \checkmark &            &            &            &            &            &            &            &            &            &         \\
                        & C3                         & \checkmark & \checkmark & \checkmark &            &            &            &            &            &            &            &         \\ \hline
\end{tabular}
\label{data_aug}%
\end{table*}

\begin{table*}[!htb]
\centering
\caption{Summary of model architecture, loss function and optimizer from top three algorithms in each task of the challenge. Abbreviation: Dice Loss (DL), Weighted Dice Loss (WDL), Focal Loss (FL), Cross Entropy loss (CE), Stochastic Gradient Descent (SGD).}
\begin{tabular}{cclll}
\hline
\multicolumn{1}{l}{}    & Algorithm & Model architecture                      & Loss function & Optimizer \\ \hline
\multirow{3}{*}{Task 1} & A1        & U2Net                                   & WDL + FL/CE   & AdamW     \\
                        & A2        & ConvNext + SegFormer + Swin Transformer & DL + FL/CE    & AdamW     \\
                        & A3        & nnUNet                                  & DL + CE       & SGD       \\ \hline
\multirow{3}{*}{Task 2} & B1        & EfficientNet                            & Smooth L1     & AdamW     \\
                        & B2        & BEIT + NFNet                            & CE            & AdamW     \\
                        & B3        & Inception-V3 + SE-ResNeXt + ViT         & CE            & SGD       \\ \hline
\multirow{3}{*}{Task 3} & C1        & EfficientNet                            & Smooth L1     & AdamW     \\
                        & C2        & BEIT                                    & CE            & AdamW     \\
                        & C3        & DenseNet121 + Efficientnet           & CE            & Adam      \\ \hline
\end{tabular}
\label{model}%
\end{table*}%

\subsection{Data pre-processing and augmentation}
Among the data pre-processing techniques, image normalization was commonly used by the participating teams. In addition, Contrast-limited Adaptive Histogram Equalization (CLAHE), bilateral filter, and colorization were also used by some of the teams.

Data augmentation is used to prevent overfitting and improve the generalization of deep learning models. In the challenge, all teams used some form of data augmentation to enhance their models. Table \ref{data_aug} summarizes the data augmentation techniques by the top three teams of each task. The first form of data augmentation is geometric transformation which is used to increase the size of the training image, such as flipping, rotation, cropping and scaling. Among them, flipping, scaling and rotation are the most commonly used techniques by participating teams. In addition, the transformations of affine and grid distortion are also used by one team, but they are not commonly used in model development by other participating teams. The second form of data augmentation is intensity transformation which changes the pixel value and appearance of the image. The popular intensity transformation techniques are brightness, contrast, gamma, sharpen and blur, while the less used ones included coarse dropout, Gaussian noise and cut out. The geometric transformation and intensity transformation are the single-image augmentation that performs the transformation and generates a new training data from one image. Unlike single-image augmentation, multi-image augmentation can generate a new training data from multiple images, which is becoming increasingly popular in model development. In particular, B3 used the multi-image augmentation techniques of MixUp \citep{zhang2017mixup} and CutMix \citep{yun2019cutmix} to enhance their model.

The above data augmentation techniques are mainly used for training data during the training phase. In testing phase, the test-time augmentation (TTA) was used by some teams for the testing data to improve the test performance. Specifically, the model first sequentially takes multiple augmented versions of the same inference image as input and produces multiple predictions, then these predictions are combined to obtain the final prediction, usually by averaging them. TTA was integrated into the first-ranked algorithms in all three tasks, and has been shown to be effective in improving the final test performance.

\subsection{Model architectures}
In the three tasks of the challenge, all teams used deep learning methods to construct the segmentation and classification frameworks. Table \ref{model} summarizes the model architectures of the top three solutions in each task. Among them, five solutions used a single network for model development, while four solutions used an ensemble of two or more networks. Among single-network solutions, U2Net \citep{qin2020u2} and nnUNet \citep{isensee2021nnu} were adopted as the model architecture for the segmentation task respectively. EfficientNet \citep{tan2019efficientnet} and BEIT \citep{bao2021beit} were adopted as the model architecture for the two classification tasks, where B1 and C1 both used EfficientNet as the network architecture. Among multi-network ensemble solutions, A2 used an ensemble of ConvNext \citep{liu2022convnet}, SegFormer \citep{xie2021segformer} and Swin Transformer \citep{liu2021swin} to develop the segmentation algorithm. In the two classification tasks, multiple forms of the ensemble of different networks were adopted, including an ensemble of BEIT \citep{bao2021beit} and NFNet \citep{brock2021high}, an ensemble of Inception-V3 \citep{szegedy2016rethinking}, SE-ResNeXt \citep{xie2017aggregated} and ViT \citep{dosovitskiy2020image}, and an ensemble of DenseNet121 \citep{huang2017densely} and Efficientnet \citep{tan2019efficientnet}.

\subsection{Model Optimization}
The choice of loss function depends on the desired output being predicted. There is a degree of variability in the loss functions between segmentation task and classification task. The summary of the loss functions used by top three teams is shown in Table \ref{model}. Among the various loss functions in segmentation task, dice loss, pixel-wise cross entropy loss and focal loss \citep{lin2017focal} were the most popular loss functions adopted by these teams. These loss functions were usually not used individually, but combined with each other, such as combination of dice loss and cross entropy loss, and combination of dice loss and focal loss. In the two classification tasks, the regular classification loss functions were used, including smooth L1 loss and cross entropy loss, with no combinations. Among various optimizers, as shown in Fig. \ref{model}, AdamW \citep{loshchilov2017decoupled} was the most commonly used by these teams in both segmentation and classification tasks. In addition, Stochastic Gradient Descent (SGD) and Adam \citep{kingma2014adam} were also presented in some of the solutions.

\begin{figure*}[htb]
	\begin{center}
		\includegraphics[width=6.5in]{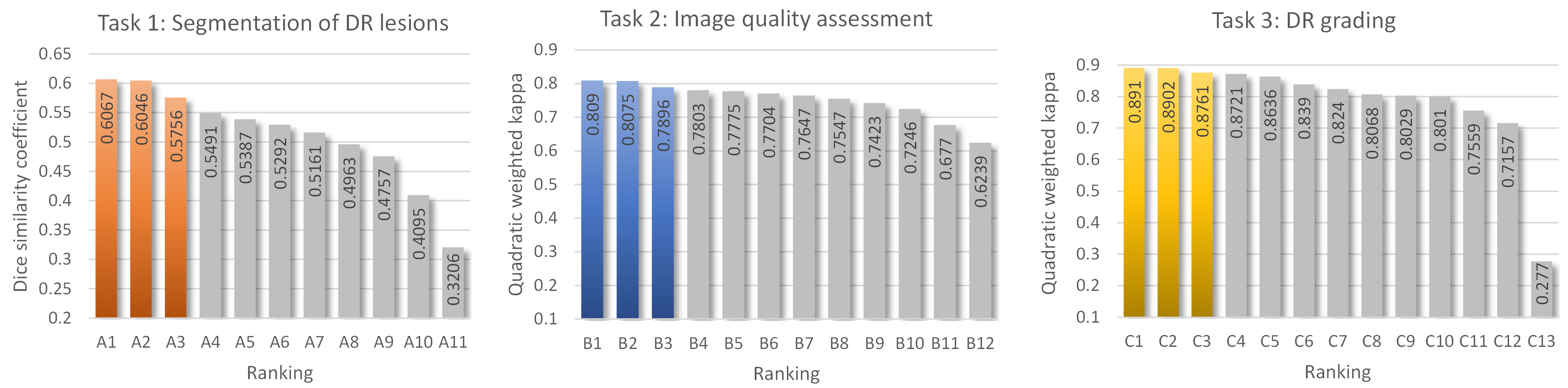}
		\caption{Bar charts of the final rankings for three tasks. The colored bars indicate the top three teams in each task.}
		\label{rank}
	\end{center}
\end{figure*}

\begin{figure*}[htb]
	\begin{center}
		\includegraphics[width=6.5in]{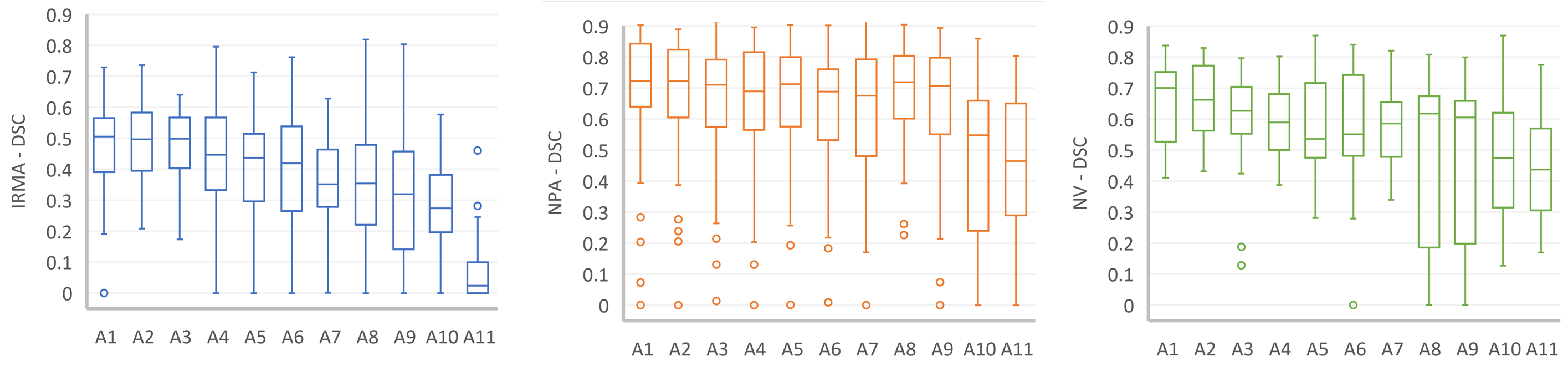}
		\caption{Box plots illustrating the DSC performance of the teams across the three lesions in DR lesion segmentation task. Abbreviation: intraretinal microvascular abnormality (IRMA), nonperfusion area (NPA), neovascularization (NV).}
		\label{task_1}
	\end{center}
\end{figure*}

\subsection{Post-processing}
Post-processing techniques are used to process the output of the deep learning model and produce the final prediction result. Among the post-processing techniques of the segmentation task, A1 used the dilation operation for the predicted NPA masks. For predicted IRMA and NV masks, if one pixel is predicted to be both IRMA and NV, then the class with high confidence will be retained. A2 used morphological logical OR operation to obtain the final mask from the multiple predicted masks by multiple networks, and the final mask of NPA is obtained in the same way but with logical AND operation. A3 also adopted the combination of multiple predicted masks to get the final prediction, but these different masks are from the same model with different configurations, such as different data augmentation techniques and dropout factors. Moreover, A3 used a union of multiple predicted masks to generate the final mask for IRMA, and used a majority voting strategy to generate the final mask for NPA and NV. In the image quality assessment task, B1 took the multiple-class classification task as a regression problem and used the class-specific operating threshold to confirm the class of the image instead of the rounding of predicted probability. In DR grading task, C1 and C2 designed the rules to combine the classification results of DR grade with lesion segmentation results to generate the final DR grade for each image, in which the lesion segmentation results can be used to correct the predicted DR grade because lesions indicate the different DR grades, such as NV lesion corresponds to PDR grade.

\section{Results}
This section presents the results obtained by the participating teams on the test set, and analyzes the ranking stability of these algorithms. We also present the ensemble of results from top three algorithms in each task, and report the statistical significance analysis of the algorithms. The final rankings of the three tasks are shown in Fig. \ref{rank}.

\subsection{Task 1: segmentation of DR lesions}
In the task of DR lesion segmentation, out of the total 24 teams that submitted the results on the test set, 11 teams submitted the method description papers. The average DSC of the top ten teams ranged from 40.95\% to 60.67\%. The DSC distribution of each class of lesion is shown in Fig. \ref{task_1}. Among these three lesions, IRMA has the lowest segmentation performance, with DSC ranging from 29.53\% to 47.04\% for the top ten algorithms. NPA has the highest segmentation performance with DSC ranging from 46.59\% to 69.26\% for the top ten algorithms, and NV ranks second with DSC ranging from 46.73\% to 65.71\% for the top ten algorithms. The complex lesion features could explain the low segmentation performance of IRMA because IRMAs are usually thin vessels that spread throughout the image.

\begin{figure*}[!htb]
	\begin{center}
		\includegraphics[width=6.5in]{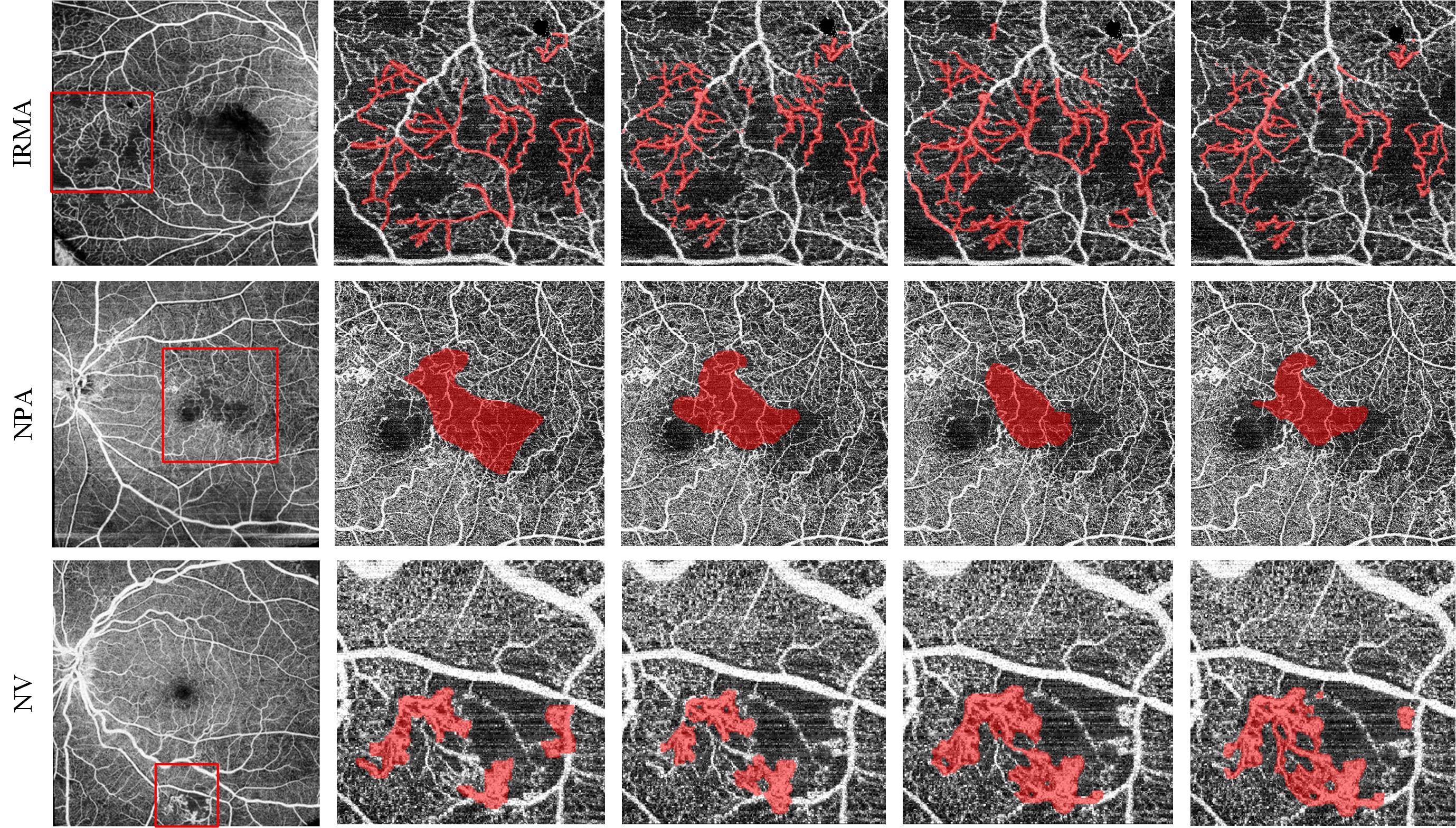}
		\caption{Visual segmentation results of top three methods in DR lesion segmentation task. From left to right in the columns are the original UW-OCTA images, ground truths, the results of the first ranked method (A1), the result of the second ranked method (A2), and the result of the third ranked method (A3), respectively.}
		\label{seg_case}
	\end{center}
\end{figure*}

In addition to the DSC and IoU used in the challenge, we also use sensitivity (SEN), precision (PRE) and specificity (SPE) to evaluate the segmentation performance of each method. The quantitative results of the top three methods are shown in Table \ref{T1_metric}, and the visual segmentation results are shown in Fig. \ref{seg_case}. For these three methods, A1 has 8 evaluation metrics ranked first, while A2 and A3 have 5 and 2 respectively. In addition to DSC, IRMA also gets the lowest scores in several other metrics, including IoU, sensitivity and precision. Conversely, IRMA obtains the highest specificity score that measures the ability of the method to identify background pixels. The highest DSC score of IRMA is achieved by A2 with a DSC of 48.32\%, and the highest DSC score of NPA and NV are both achieved by A1, with a DSC of 69.26\% and 65.71\% respectively.

\begin{table*}[!htb]
\small
\centering
\caption{Quantitative results of top three teams in DR lesion segmentation task. Ensemble represents the ensemble results of the top three algorithms with a majority voting strategy. The bold denotes the best score within the top three algorithms.}
\begin{tabular}{cccccccccccccccc}
\hline
\multirow{2}{*}{Algorithm} & \multicolumn{5}{c}{IRMA}                                                           & \multicolumn{5}{c}{NPA}                                                   & \multicolumn{5}{c}{NV}                                                             \\
                      & DSC            & IoU            & SEN            & PRE            & SPE            & DSC            & IoU            & SEN            & PRE            & SPE   & DSC            & IoU            & SEN            & PRE            & SPE            \\ \hline
A1                     & 47.04          & 31.89          & 53.01          & \textbf{47.54} & 99.36          & \textbf{69.26} & \textbf{55.66} & 73.57          & \textbf{71.00} & \textbf{95.75} & \textbf{65.71} & 50.15          & 73.35          & \textbf{62.51} & \textbf{98.93} \\
A2                     & \textbf{48.32} & \textbf{32.99} & \textbf{57.56} & 47.07          & 99.19          & 67.36          & 53.96          & 74.72          & 69.20          & 94.47 & 65.70          & \textbf{50.24} & \textbf{77.07} & 60.33          & 98.68          \\
A3                     & 46.72          & 31.45          & 57.37          & 43.21          & \textbf{99.40} & 66.80          & 52.60          & \textbf{75.24} & 67.14          & 94.82 & 59.17          & 44.33          & 64.27          & 60.80          & 99.30          \\ \hline
Ensemble                     & 49.95          & 34.35          & 57.00          & 49.74          & 99.40          & 68.78          & 55.26          & 74.94          & 70.25          & 95.09 & 67.55          & 52.16          & 73.13          & 65.53          & 99.20          \\ \hline
\end{tabular}
\label{T1_metric}
\end{table*}

\subsection{Task 2: image quality assessment}
In the task of image quality assessment, a total of 45 teams submitted the results, and 12 teams submitted the method description papers where the quadratic weighted kappa of the top ten teams ranged from 0.7246 to 0.8090. Table \ref{iql} shows the quantitative results of the top three algorithms for the metrics of sensitivity, specificity and F1 score. Combined with the confusion matrices of the top three methods in Fig. \ref{CM}, we can see that misclassification usually occurs between poor and good image quality levels. For each of these three methods, there are many images that have a poor quality level but are predicted to be in good quality level, while many images with a good quality level are predicted to be in excellent quality level. There are mainly two reasons to explain the results. One is the data imbalance of the dataset in which 80\% of the images have an excellent image quality level, and the rest of 20\% have a poor and good quality level. Another is the small feature differences between different image quality levels, especially between two adjacent image quality levels, making it difficult for the network to classify correctly.

\subsection{Task 3: DR grading}
\begin{figure*}[htb]
	\begin{center}
		\includegraphics[width=6.5in]{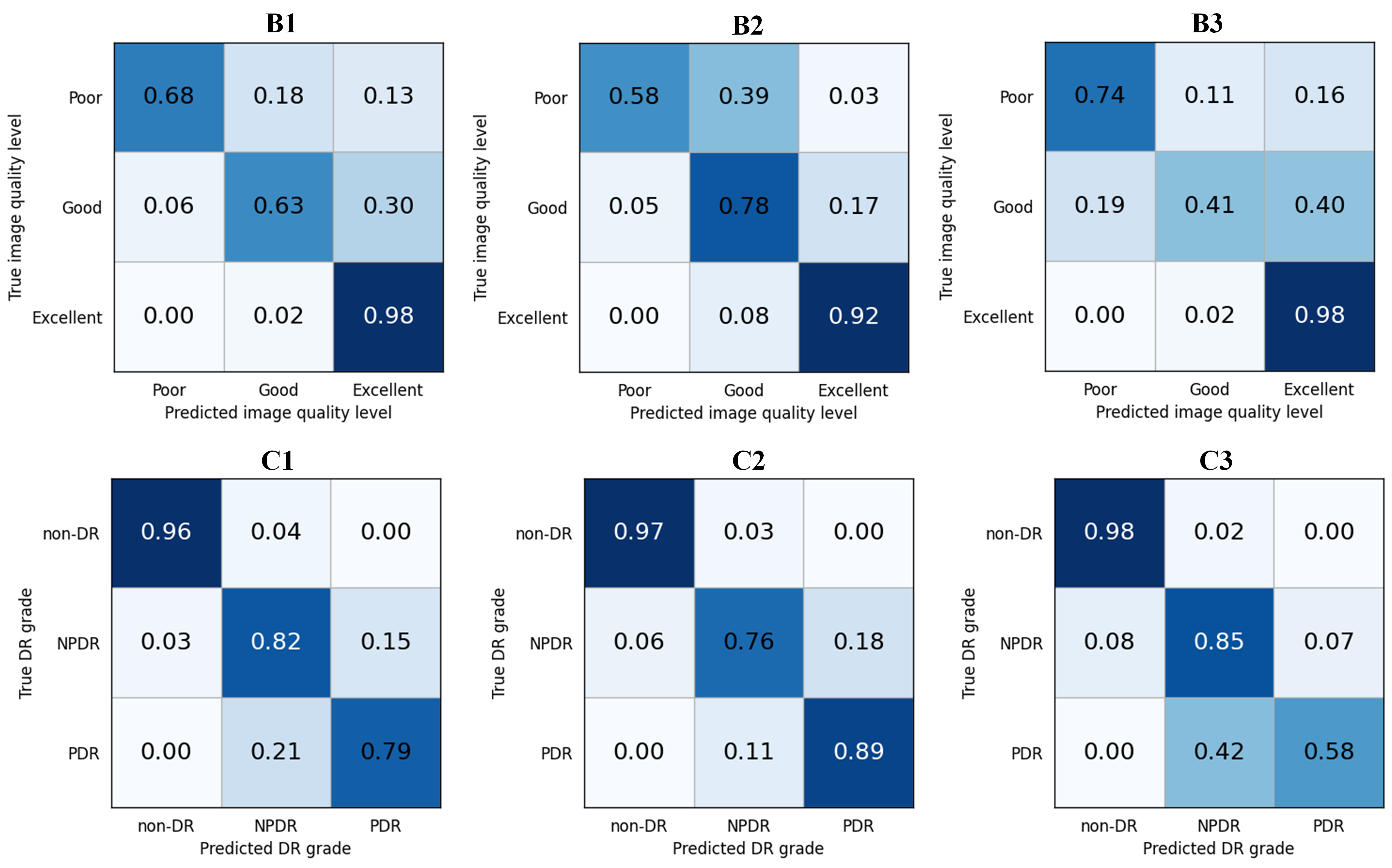}
		\caption{Confusion matrix for image quality assessment task (top row) and DR grading task (bottom row). From left to right are the results of the first, second and third ranked algorithms, respectively.}
		\label{CM}
	\end{center}
\end{figure*}

In the task of DR grading, a total of 45 teams submitted the results, and 13 teams submitted the method description papers where the quadratic weighted kappa of the top ten teams ranged from 0.7157 to 0.8910. Table \ref{dr} shows the quantitative results of the top three algorithms for the metrics of sensitivity, specificity and F1 score. Combined with the confusion matrices of the top three methods in Fig. \ref{CM}, we can see that misclassification usually occurs between NPDR and PDR. One of the reasons is the data imbalance where the non-DR, NPDR and PDR occupy about 55\%, 35\% and 10\% of the images in the dataset respectively, and another reason is the small feature differences between these two classes, which makes it difficult for the network to extract effective discriminative features of these two classes.

\begin{table}[!ht]
\small
\centering
\caption{Quantitative results of the top three algorithms in image quality assessment task. Ensemble represents the ensemble results of the top three algorithms with a majority voting strategy.}
\begin{tabular}{ccccc}
\hline
Algorithm                 & class     & Sensitivity & Specificity & F1 score \\ \hline
\multirow{3}{*}{B1}       & Poor      & 0.6842      & 0.9875      & 0.7536   \\
                          & Good      & 0.6349      & 0.9627      & 0.6838   \\
                          & Excellent & 0.9763      & 0.7624      & 0.9536   \\
\multirow{3}{*}{B2}       & Poor      & 0.5789      & 0.9925      & 0.6984   \\
                          & Good      & 0.7778      & 0.8853      & 0.6323   \\
                          & Excellent & 0.9169      & 0.8812      & 0.9392   \\
\multirow{3}{*}{B3}       & Poor      & 0.7368      & 0.9700      & 0.7179   \\
                          & Good      & 0.4127      & 0.9733      & 0.5253   \\
                          & Excellent & 0.9822      & 0.6931      & 0.9471   \\ \hline
\multirow{3}{*}{Ensemble} & Poor      & 0.6842      & 0.9875      & 0.7536   \\
                          & Good      & 0.6349      & 0.9627      & 0.6838   \\
                          & Excellent & 0.9763      & 0.7624      & 0.9536   \\ \hline
\end{tabular}
\label{iql}
\end{table}

\begin{table}[!ht]
\small
\centering
\caption{Quantitative results of top three teams in DR grading task. Ensemble represents the ensemble results of the top three algorithms with a majority voting strategy.}
\begin{tabular}{ccccc}
\hline
Algorithm                 & class  & Sensitivity & Specificity & F1 score \\ \hline
\multirow{3}{*}{C1}       & non-DR & 0.9631      & 0.9763      & 0.9721   \\
                          & NPDR   & 0.8168      & 0.9373      & 0.8425   \\
                          & PDR    & 0.7895      & 0.9425      & 0.6818   \\
\multirow{3}{*}{C2}       & non-DR & 0.9724      & 0.9527      & 0.9679   \\
                          & NPDR   & 0.7557      & 0.9608      & 0.8250    \\
                          & PDR    & 0.8947      & 0.9310       & 0.7083   \\
\multirow{3}{*}{C3}       & non-DR & 0.9770      & 0.9349      & 0.9636   \\
                          & NPDR   & 0.8473      & 0.9176      & 0.8441   \\
                          & PDR    & 0.5789      & 0.9741      & 0.6377   \\ \hline
\multirow{3}{*}{Ensemble} & non-DR & 0.9724      & 0.9645      & 0.9724   \\
                          & NPDR   & 0.8473      & 0.9451      & 0.8672   \\
                          & PDR    & 0.7895      & 0.9598      & 0.7317   \\ \hline
\end{tabular}
\label{dr}
\end{table}

\subsection{Ranking stability}
\begin{figure}[htb]
	\begin{center}
		\includegraphics[width=3.4in]{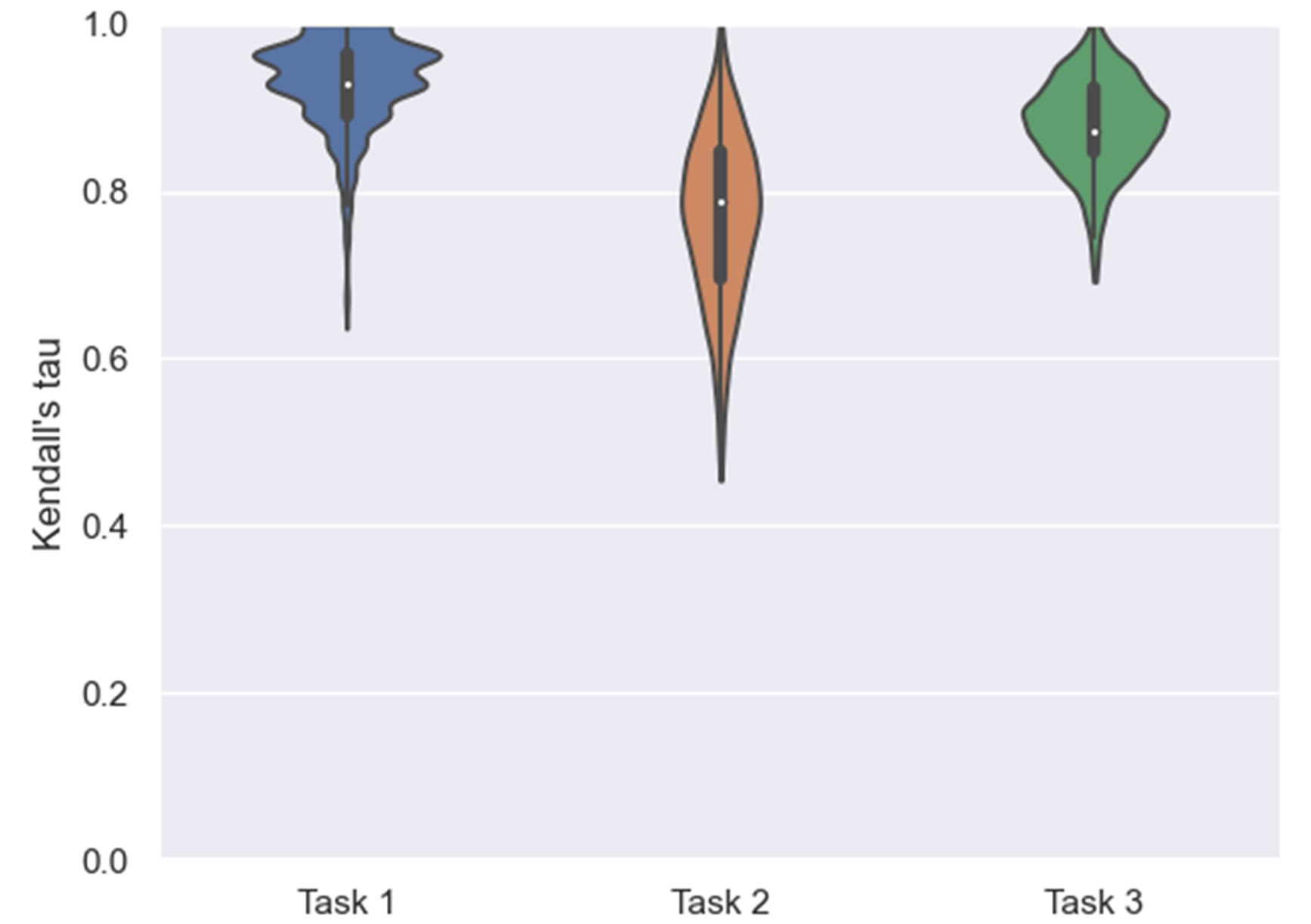}
		\caption{Violin plots of Kendall's $\tau$ for visualizing ranking stability based on bootstrapping. 1000 bootstrap samples are used for each task. The ranking list based on the full test data is compared pairwise with the ranking lists based on the individual bootstrap samples. Kendall’s $\tau$ is computed for each pair of rankings, and a violin plot that simultaneously depicts a boxplot and a density plot is generated from the results.}
		\label{kendall}
	\end{center}
\end{figure}

\begin{figure*}[t]
	\begin{center}
		\includegraphics[width=6.5in]{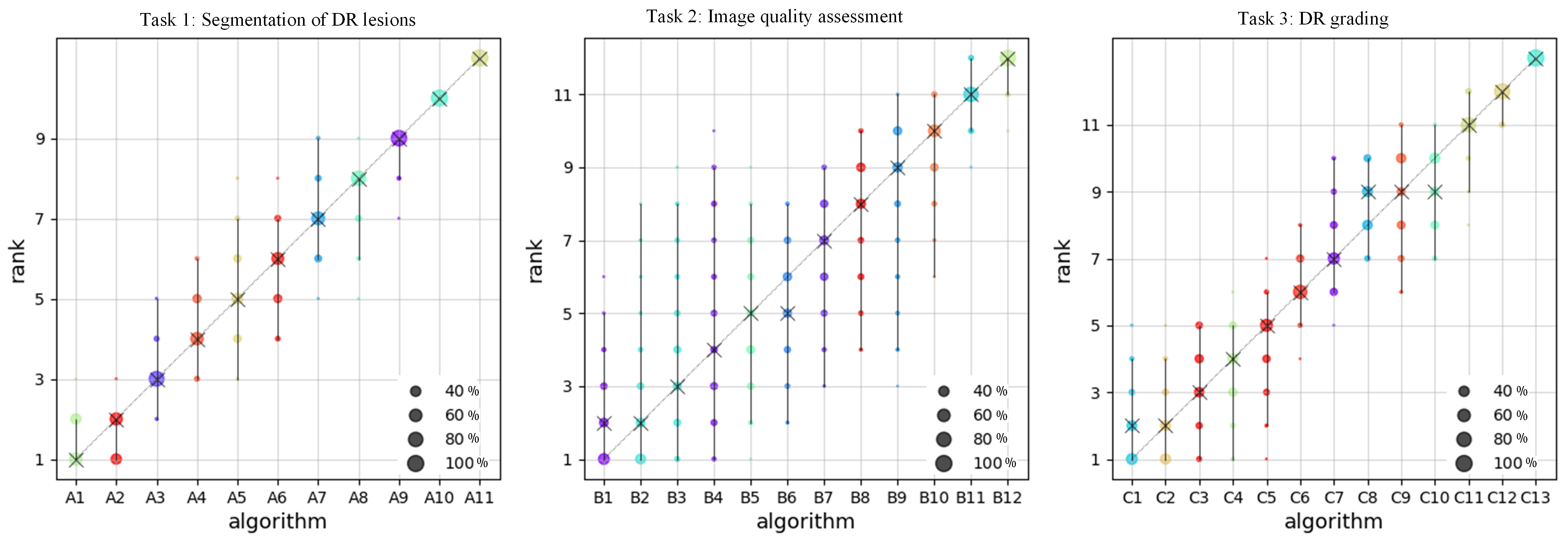}
		\caption{Blob plots for visualizing ranking stability based on bootstrap sampling. 1000 bootstrap samples are used for each task. The size of each circle is proportional to the relative frequency an algorithm obtained the corresponding rank across 1000 bootstrap samples. The median rank for each algorithm is indicated by a black cross. 95\% bootstrap intervals across bootstrap samples are indicated by black lines.}
		\label{ranking_stab}
	\end{center}
\end{figure*}

Inspired by the challengeR toolkit \citep{wiesenfarth2021methods}, we performed bootstrapping (1000 bootstrap samples) to assess the stability of rankings with respect to sampling variability. To quantitatively assess the ranking stability, the agreement of the challenge ranking and the ranking of each bootstrap on the test set was determined via Kendall’s $\tau$, which is a rank correlation coefficient with a value between -1 (reverse ranking order) and 1 (identical ranking order). The Violin plots shown in Fig. \ref{kendall} illustrate the bootstrap results for each task. We obtained a Kendall’s $\tau$ of 0.9313, 0.7697 and 0.8802 for the tasks of DR lesion segmentation, image quality assessment and DR grading respectively. Fig. \ref{ranking_stab} shows a blob plot of the bootstrap rankings for each task.

\begin{figure*}[htb]
	\begin{center}
		\includegraphics[width=6.5in]{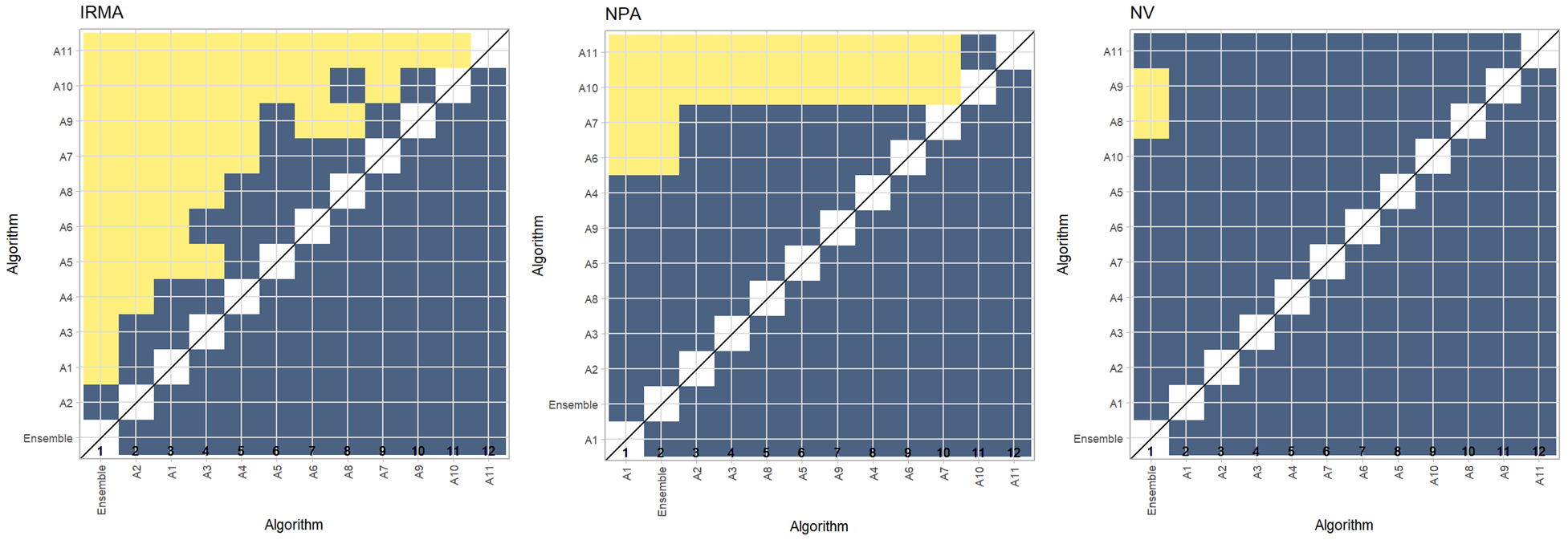}
		\caption{Significance maps depict incidence matrices of pairwise significant test results in DR lesion segmentation task for the one-sided Wilcoxon signed rank test at a 5\% significance level with adjustment for multiple testing according to Holm. Yellow shading indicates that the mean DSC of the algorithm on the x-axis is significantly superior to those from the algorithm on the y-axis, and blue color indicates no significant difference.}
		\label{significance}
	\end{center}
\end{figure*}

\subsection{Ensemble of top three algorithms}
The ensemble of networks has shown great power in improving model performance \citep{ganaie2021ensemble, xie2016melanoma, khened2019fully}. Thus it is interesting to explore the ensemble of the top three methods in each task. For the segmentation task, we use three forms of ensemble to generate the final segmentation output, including logical AND, logical OR and majority voting. The DSC of the three forms of ensemble results are 47.10\%, 45.36\% and 49.95\% for IRMA, 67.91\%, 66.83\% and 68.78\% for NPA, and 63.33\%, 60.23\% and 67.55\% for NV. For each of the three classes, the highest DSC is achieved by the majority voting strategy, of which the averaged DSC is 62.09\% which is 1.42\% higher than the best result of the participating algorithms. The detailed performance of the ensemble result with a majority voting is shown in Table \ref{T1_metric}. For the two classification tasks, we use a majority voting strategy from the top three results to generate the ensemble result. For image quality assessment, the quadratic weighted kappa of the ensemble result is 0.8090, which is equal to the best score of participating algorithms. The detailed performance of the ensemble result is shown in Table \ref{iql}. For DR grading, the quadratic weighted kappa of the ensemble result is 0.9044, which is 1.4\% higher than the best score of participating algorithms. The detailed performance of the ensemble result is shown in Table \ref{dr}. From the ensemble results in both the segmentation task and classification task, we can see that the ensemble of the model has great power in improving the performance of deep learning methods.

\subsection{Statistical significance analysis of the algorithms}
\begin{figure}[!htb]
	\begin{center}
		\includegraphics[width=3.4in]{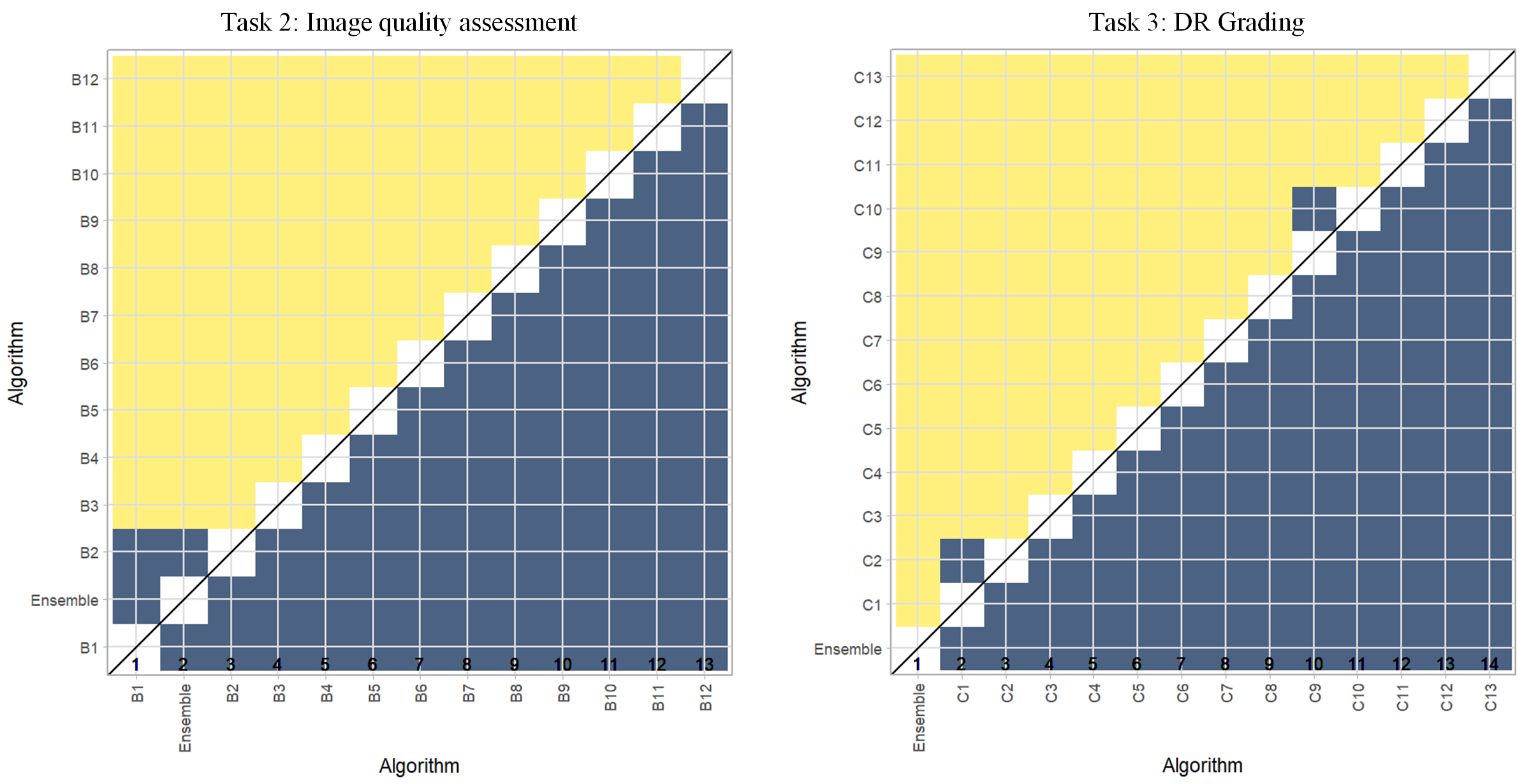}
		\caption{Significance maps depict incidence matrices of pairwise significant test results for the one-sided Wilcoxon signed rank test at a 5\% significance level with adjustment for multiple testing according to Holm. Yellow shading indicates that the quadratic weighted kappa from the algorithm on the x-axis is significantly superior to those from the algorithm on the y-axis, and blue color indicates no significant difference.}
		\label{significance_cls}
	\end{center}
\end{figure}

For each task of challenge, the statistical comparison of the score of each team is done with the one-tailed Wilcoxon signed rank test at a 5\% significance level. The challengeR toolkit \citep{wiesenfarth2021methods} is used to perform the significance analysis and generate the significance map. For the task of DR lesion segmentation, the significance maps in Fig. \ref{significance} show the results of the statistical significance analysis for the three lesions. In conclusion, there is a significant difference between multiple algorithm pairs for IRMA segmentation, but for the segmentation of NPA and NV, only a few pairs of algorithms show significant differences. For the two classification tasks, the statistical comparison of the score of each team is shown in Fig. \ref{significance_cls}. In the task of image quality assessment, there are no significant differences between the algorithms B1, ensemble model and B2. In the task of DR grading, the ensemble model is significantly superior to the algorithms C1, C2 and C3, but there is no significant difference between C1 and C2.

\section{Discussion}
In this paper, we present the details of the DRAC challenge, including the challenge setup, dataset, and evaluation, as well as a comprehensive analysis of competing solutions and results. In each task of the challenge, there are more than 10 teams that submitted the method description papers reporting their solutions and results. These solutions have given detailed research on data augmentation, model architecture and post-processing techniques, advancing the collective knowledge of the research community for medical image analysis.

\subsection{Analysis of the methods}
All teams used deep learning methods to solve the segmentation and classification problems in this challenge, demonstrating the great potential and effectiveness of deep learning. First of all, all teams used some forms of data augmentation techniques to increase the training size and reduce overfitting. By varying the types and parameters of image transformations, a large amount of new training images can be generated to enhance the generalization of the models, especially for medical image datasets that usually contain fewer images. The most popular image transformation is flipping, which was adopted by almost all top-performing teams. Other commonly used image transformations include rotation, scaling, brightness modification and contrast modification, etc. In addition, more complex multi-image data augmentation techniques are also effective for improving model performance, including MixUp and CutMix. It is possible that more of these data augmentation techniques will emerge in the future. However, the challenge results clearly show that the types of data augmentation used are not as much as possible and as complex as possible. Which data augmentation techniques are used depends on the specific dataset and model architecture, and needs to be further verified through experiments.

Model architecture is the most important part of developing a solution for segmentation and classification tasks. Some teams used a single network for model development, while some teams used an ensemble of multiple networks. The variants of UNet, such as U2Net and nnUNet, performed well in the segmentation task. The currently popular ConvNext and transformer-based segmentation networks have also achieved good results. In the classification task, EfficientNet, BEIT, and DenseNet showed great power and received much attention from participating teams. Through analyzing the methods from the top ranked solutions, it is clear that the combination of multiple predictions to generate the final prediction has become an important trend in both segmentation and classification tasks. The multiple predictions can be from a single network with multiple transformed images of the same inference image as input, or from multiple networks, or from one same network architecture but with different training configurations. The popular ways of the combination of these predictions are logical OR, logical AND and the majority voting strategy. This multi-prediction combination is similar to the consistency-based learning that is used in the semi-supervision learning methods \citep{ouali2020semi, bortsova2019semi}. Another interesting technique that was used in image quality assessment and DR grading tasks involves formulating the multi-class classification problem as a regression problem in order to consider the correlation among classes, because these classes correspond to different degrees of severity of the disease. Lastly, the presented solutions also show diversity in loss function between segmentation and classification tasks. The segmentation task can be considered as a pixel-wise classification task but is more complex. The loss used in the segmentation task is usually a combination of multiple loss functions, where the commonly used are dice loss, cross entropy loss and focal loss. Also, the results showed that the loss functions that are sensitive to metrics, such as dice loss and jaccard loss, are effective for improving the score of the corresponding metrics \citep{bertels2019optimizing}.

\subsection{UW-OCTA images for DR analysis}
In clinical diagnosis, multiple imaging modalities are used for the analysis of DR, including fundus photography and optical coherence tomography angiography (OCTA). Fundus photography is the most commonly used modality for rapid screening of DR. There are already some challenges that provide benchmarks for the evaluation of fundus photography for DR analysis, such as DeepDRiD \citep{liu2022deepdrid} and IDRiD \citep{porwal2020idrid}. A large amount of state-of-the-art algorithms are developed for these challenge tasks and contribute a lot to the research community. In recent years, as a non-invasive, safe and easily repeatable retinal blood flow imaging modality, ultra-wide OCTA (UW-OCTA) has attracted much attention from researchers in diagnosis, screening, and follow-up of DR \citep{khalid2021widefield, cui2021comparison, pichi2020wide}. The UW-OCTA images may have a higher or comparable detection rate for some DR lesions or DR grades than fundus photography. Therefore, the advanced deep learning algorithms have flexible applications for DR analysis using UW-OCTA images. However, as far as we know, there is a lack of publicly available UW-OCTA datasets for fair performance evaluation of the different algorithms. Against this background, we organize the Diabetic Retinopathy Analysis Challenge, aiming to provide a benchmark and evaluation framework for automatic diabetic retinopathy analysis from UW-OCTA images. In this challenge, multiple novel deep learning algorithms have been proposed for the segmentation of DR lesions, image quality assessment and DR grading, which is of great value to the DR research community.

\subsection{Limitations of the study}
In this challenge, we provided the training data and testing data for the training and evaluation of three DR-related tasks. One limitation is the data imbalance within the different classes in the dataset, which will cause the model to have better performance for a class that contains more images and lower performance for a class that contains fewer images. Although data augmentation can be used to mitigate this data imbalance, it cannot completely eliminate this imbalance. For instance, in the image quality assessment task, the images with excellent quality levels occupy about 80\% of the dataset, so this class has the highest 98\% of accuracy which is about 20\% higher than the other two classes. The same happens in the other two tasks. Another limitation is the limited number of images in DR lesion segmentation task, especially for NV segmentation. In both the training (109 images) and test sets (65 images), only about 25\% of the images contained NV lesions. Therefore more data is needed for model development and evaluation.

\subsection{Future work}
Besides the methods and results presented in this paper, we also made the DRAC dataset available for the community, which can be downloaded from the DRAC website. We hope the dataset will be of high value for the community to address some research topics in the field, such as segmentation of DR lesions, image quality assessment and DR grading. In addition, we will keep the post-challenge registrations and submissions open, hoping more novel solutions can be evaluated and drive forward innovation for automatic diabetic retinopathy analysis. By the end of January 18, 2023, there have been 12, 10 and 12 post-challenge submissions for tasks of DR lesion segmentation, image quality assessment and DR grading respectively.

\section{Conclusion}
The DRAC challenge hosted at MICCAI 2022 provides a benchmark and evaluation framework for automatic diabetic retinopathy analysis from UW-OCTA images, including the tasks of DR lesion segmentation, image quality assessment and DR grading. With numerous participants from academia and industry around the world, the challenge provides various solutions and comparable diagnosis results for ophthalmologists on DR analysis. We thoroughly presented and discussed the algorithms and results from the participating teams in the challenge. These algorithms are likely to be integrated into the computer-aided automatic diagnostic system for diabetic retinopathy in the future, which can help reduce the burden on healthcare workers and improve the accuracy of DR diagnosis. Nevertheless, further research is still needed to improve the model and achieve a clinically available diagnostic approach for DR. To date, the challenge website remains open for post-challenge registrations and submissions, aiming to provide a sustainable benchmark and evaluation platform for the research community.

\bibliography{refs}

\begin{thebibliography}{10}
\urlstyle{rm}
\expandafter\ifx\csname url\endcsname\relax
  \def\url#1{\texttt{#1}}\fi
\expandafter\ifx\csname urlprefix\endcsname\relax\def\urlprefix{URL }\fi
\expandafter\ifx\csname doiprefix\endcsname\relax\def\doiprefix{DOI: }\fi
\providecommand{\bibinfo}[2]{#2}
\providecommand{\eprint}[2][]{\url{#2}}

\bibitem{reichel2015diabetic}
\bibinfo{author}{Reichel, E.} \& \bibinfo{author}{Salz, D.}
\newblock \bibinfo{title}{Diabetic retinopathy screening}.
\newblock In \emph{\bibinfo{booktitle}{Managing Diabetic Eye Disease in
  Clinical Practice}}, \bibinfo{pages}{25--38} (\bibinfo{publisher}{Springer},
  \bibinfo{year}{2015}).

\bibitem{atlas2017brussels}
\bibinfo{author}{Atlas, I.~D.}
\newblock \bibinfo{journal}{\bibinfo{title}{Brussels, belgium: international
  diabetes federation; 2013}}.
\newblock {\emph{\JournalTitle{International Diabetes Federation (IDF)}}}
  \textbf{\bibinfo{volume}{147}} (\bibinfo{year}{2017}).

\bibitem{wang2018diabetic}
\bibinfo{author}{Wang, W.} \& \bibinfo{author}{Lo, A.~C.}
\newblock \bibinfo{journal}{\bibinfo{title}{Diabetic retinopathy:
  pathophysiology and treatments}}.
\newblock {\emph{\JournalTitle{International journal of molecular sciences}}}
  \textbf{\bibinfo{volume}{19}}, \bibinfo{pages}{1816} (\bibinfo{year}{2018}).

\bibitem{american202011}
\bibinfo{author}{Association, A.~D.}
\newblock \bibinfo{journal}{\bibinfo{title}{11. microvascular complications and
  foot care: standards of medical care in diabetes- 2020}}.
\newblock {\emph{\JournalTitle{Diabetes care}}} \textbf{\bibinfo{volume}{43}},
  \bibinfo{pages}{S135--S151} (\bibinfo{year}{2020}).

\bibitem{jones2010diabetic}
\bibinfo{author}{Jones, S.} \& \bibinfo{author}{Edwards, R.}
\newblock \bibinfo{journal}{\bibinfo{title}{Diabetic retinopathy screening: a
  systematic review of the economic evidence}}.
\newblock {\emph{\JournalTitle{Diabetic medicine}}}
  \textbf{\bibinfo{volume}{27}}, \bibinfo{pages}{249--256}
  (\bibinfo{year}{2010}).

\bibitem{lin2016addressing}
\bibinfo{author}{Lin, S.}, \bibinfo{author}{Ramulu, P.},
  \bibinfo{author}{Lamoureux, E.~L.} \& \bibinfo{author}{Sabanayagam, C.}
\newblock \bibinfo{journal}{\bibinfo{title}{Addressing risk factors, screening,
  and preventative treatment for diabetic retinopathy in developing countries:
  a review}}.
\newblock {\emph{\JournalTitle{Clinical \& experimental ophthalmology}}}
  \textbf{\bibinfo{volume}{44}}, \bibinfo{pages}{300--320}
  (\bibinfo{year}{2016}).

\bibitem{wang2017availability}
\bibinfo{author}{Wang, L.~Z.} \emph{et~al.}
\newblock \bibinfo{journal}{\bibinfo{title}{Availability and variability in
  guidelines on diabetic retinopathy screening in asian countries}}.
\newblock {\emph{\JournalTitle{British Journal of Ophthalmology}}}
  \textbf{\bibinfo{volume}{101}}, \bibinfo{pages}{1352--1360}
  (\bibinfo{year}{2017}).

\bibitem{ting2016diabetic}
\bibinfo{author}{Ting, D. S.~W.}, \bibinfo{author}{Cheung, G. C.~M.} \&
  \bibinfo{author}{Wong, T.~Y.}
\newblock \bibinfo{journal}{\bibinfo{title}{Diabetic retinopathy: global
  prevalence, major risk factors, screening practices and public health
  challenges: a review}}.
\newblock {\emph{\JournalTitle{Clinical \& experimental ophthalmology}}}
  \textbf{\bibinfo{volume}{44}}, \bibinfo{pages}{260--277}
  (\bibinfo{year}{2016}).

\bibitem{dai2021deep}
\bibinfo{author}{Dai, L.} \emph{et~al.}
\newblock \bibinfo{journal}{\bibinfo{title}{A deep learning system for
  detecting diabetic retinopathy across the disease spectrum}}.
\newblock {\emph{\JournalTitle{Nature communications}}}
  \textbf{\bibinfo{volume}{12}}, \bibinfo{pages}{1--11} (\bibinfo{year}{2021}).

\bibitem{jelinek2009automated}
\bibinfo{author}{Jelinek, H.} \& \bibinfo{author}{Cree, M.~J.}
\newblock \emph{\bibinfo{title}{Automated image detection of retinal
  pathology}} (\bibinfo{publisher}{Crc Press}, \bibinfo{year}{2009}).

\bibitem{li2018automated}
\bibinfo{author}{Li, Z.} \emph{et~al.}
\newblock \bibinfo{journal}{\bibinfo{title}{An automated grading system for
  detection of vision-threatening referable diabetic retinopathy on the basis
  of color fundus photographs}}.
\newblock {\emph{\JournalTitle{Diabetes care}}} \textbf{\bibinfo{volume}{41}},
  \bibinfo{pages}{2509--2516} (\bibinfo{year}{2018}).

\bibitem{gulshan2016development}
\bibinfo{author}{Gulshan, V.} \emph{et~al.}
\newblock \bibinfo{journal}{\bibinfo{title}{Development and validation of a
  deep learning algorithm for detection of diabetic retinopathy in retinal
  fundus photographs}}.
\newblock {\emph{\JournalTitle{Jama}}} \textbf{\bibinfo{volume}{316}},
  \bibinfo{pages}{2402--2410} (\bibinfo{year}{2016}).

\bibitem{niemeijer2009retinopathy}
\bibinfo{author}{Niemeijer, M.} \emph{et~al.}
\newblock \bibinfo{journal}{\bibinfo{title}{Retinopathy online challenge:
  automatic detection of microaneurysms in digital color fundus photographs}}.
\newblock {\emph{\JournalTitle{IEEE transactions on medical imaging}}}
  \textbf{\bibinfo{volume}{29}}, \bibinfo{pages}{185--195}
  (\bibinfo{year}{2009}).

\bibitem{porwal2020idrid}
\bibinfo{author}{Porwal, P.} \emph{et~al.}
\newblock \bibinfo{journal}{\bibinfo{title}{Idrid: Diabetic
  retinopathy--segmentation and grading challenge}}.
\newblock {\emph{\JournalTitle{Medical image analysis}}}
  \textbf{\bibinfo{volume}{59}}, \bibinfo{pages}{101561}
  (\bibinfo{year}{2020}).

\bibitem{liu2022deepdrid}
\bibinfo{author}{Liu, R.} \emph{et~al.}
\newblock \bibinfo{journal}{\bibinfo{title}{Deepdrid: Diabetic
  retinopathy—grading and image quality estimation challenge}}.
\newblock {\emph{\JournalTitle{Patterns}}} \bibinfo{pages}{100512}
  (\bibinfo{year}{2022}).

\bibitem{maier2020bias}
\bibinfo{author}{Maier-Hein, L.} \emph{et~al.}
\newblock \bibinfo{journal}{\bibinfo{title}{Bias: Transparent reporting of
  biomedical image analysis challenges}}.
\newblock {\emph{\JournalTitle{Medical image analysis}}}
  \textbf{\bibinfo{volume}{66}}, \bibinfo{pages}{101796}
  (\bibinfo{year}{2020}).

\bibitem{kawai2021prevention}
\bibinfo{author}{Kawai, K.} \emph{et~al.}
\newblock \bibinfo{journal}{\bibinfo{title}{Prevention of image quality
  degradation in wider field optical coherence tomography angiography images
  via image averaging}}.
\newblock {\emph{\JournalTitle{Translational vision science \& technology}}}
  \textbf{\bibinfo{volume}{10}}, \bibinfo{pages}{16--16}
  (\bibinfo{year}{2021}).

\bibitem{wang2021deep}
\bibinfo{author}{Wang, Y.} \emph{et~al.}
\newblock \bibinfo{journal}{\bibinfo{title}{A deep learning-based quality
  assessment and segmentation system with a large-scale benchmark dataset for
  optical coherence tomographic angiography image}}.
\newblock {\emph{\JournalTitle{arXiv preprint arXiv:2107.10476}}}
  (\bibinfo{year}{2021}).

\bibitem{world2006prevention}
\bibinfo{author}{Organization, W.~H.}
\newblock \emph{\bibinfo{title}{Prevention of blindness from diabetes mellitus:
  report of a WHO consultation in Geneva, Switzerland, 9-11 November 2005}}
  (\bibinfo{publisher}{World Health Organization}, \bibinfo{year}{2006}).

\bibitem{zhang2017mixup}
\bibinfo{author}{Zhang, H.}, \bibinfo{author}{Cisse, M.},
  \bibinfo{author}{Dauphin, Y.~N.} \& \bibinfo{author}{Lopez-Paz, D.}
\newblock \bibinfo{journal}{\bibinfo{title}{mixup: Beyond empirical risk
  minimization}}.
\newblock {\emph{\JournalTitle{arXiv preprint arXiv:1710.09412}}}
  (\bibinfo{year}{2017}).

\bibitem{yun2019cutmix}
\bibinfo{author}{Yun, S.} \emph{et~al.}
\newblock \bibinfo{title}{Cutmix: Regularization strategy to train strong
  classifiers with localizable features}.
\newblock In \emph{\bibinfo{booktitle}{Proceedings of the IEEE/CVF
  international conference on computer vision}}, \bibinfo{pages}{6023--6032}
  (\bibinfo{year}{2019}).

\bibitem{qin2020u2}
\bibinfo{author}{Qin, X.} \emph{et~al.}
\newblock \bibinfo{journal}{\bibinfo{title}{U2-net: Going deeper with nested
  u-structure for salient object detection}}.
\newblock {\emph{\JournalTitle{Pattern recognition}}}
  \textbf{\bibinfo{volume}{106}}, \bibinfo{pages}{107404}
  (\bibinfo{year}{2020}).

\bibitem{isensee2021nnu}
\bibinfo{author}{Isensee, F.}, \bibinfo{author}{Jaeger, P.~F.},
  \bibinfo{author}{Kohl, S.~A.}, \bibinfo{author}{Petersen, J.} \&
  \bibinfo{author}{Maier-Hein, K.~H.}
\newblock \bibinfo{journal}{\bibinfo{title}{nnu-net: a self-configuring method
  for deep learning-based biomedical image segmentation}}.
\newblock {\emph{\JournalTitle{Nature methods}}} \textbf{\bibinfo{volume}{18}},
  \bibinfo{pages}{203--211} (\bibinfo{year}{2021}).

\bibitem{tan2019efficientnet}
\bibinfo{author}{Tan, M.} \& \bibinfo{author}{Le, Q.}
\newblock \bibinfo{title}{Efficientnet: Rethinking model scaling for
  convolutional neural networks}.
\newblock In \emph{\bibinfo{booktitle}{International conference on machine
  learning}}, \bibinfo{pages}{6105--6114} (\bibinfo{organization}{PMLR},
  \bibinfo{year}{2019}).

\bibitem{bao2021beit}
\bibinfo{author}{Bao, H.}, \bibinfo{author}{Dong, L.} \& \bibinfo{author}{Wei,
  F.}
\newblock \bibinfo{journal}{\bibinfo{title}{Beit: Bert pre-training of image
  transformers}}.
\newblock {\emph{\JournalTitle{arXiv preprint arXiv:2106.08254}}}
  (\bibinfo{year}{2021}).

\bibitem{liu2022convnet}
\bibinfo{author}{Liu, Z.} \emph{et~al.}
\newblock \bibinfo{title}{A convnet for the 2020s}.
\newblock In \emph{\bibinfo{booktitle}{Proceedings of the IEEE/CVF Conference
  on Computer Vision and Pattern Recognition}}, \bibinfo{pages}{11976--11986}
  (\bibinfo{year}{2022}).

\bibitem{xie2021segformer}
\bibinfo{author}{Xie, E.} \emph{et~al.}
\newblock \bibinfo{journal}{\bibinfo{title}{Segformer: Simple and efficient
  design for semantic segmentation with transformers}}.
\newblock {\emph{\JournalTitle{Advances in Neural Information Processing
  Systems}}} \textbf{\bibinfo{volume}{34}}, \bibinfo{pages}{12077--12090}
  (\bibinfo{year}{2021}).

\bibitem{liu2021swin}
\bibinfo{author}{Liu, Z.} \emph{et~al.}
\newblock \bibinfo{title}{Swin transformer: Hierarchical vision transformer
  using shifted windows}.
\newblock In \emph{\bibinfo{booktitle}{Proceedings of the IEEE/CVF
  International Conference on Computer Vision}}, \bibinfo{pages}{10012--10022}
  (\bibinfo{year}{2021}).

\bibitem{brock2021high}
\bibinfo{author}{Brock, A.}, \bibinfo{author}{De, S.}, \bibinfo{author}{Smith,
  S.~L.} \& \bibinfo{author}{Simonyan, K.}
\newblock \bibinfo{title}{High-performance large-scale image recognition
  without normalization}.
\newblock In \emph{\bibinfo{booktitle}{International Conference on Machine
  Learning}}, \bibinfo{pages}{1059--1071} (\bibinfo{organization}{PMLR},
  \bibinfo{year}{2021}).

\bibitem{szegedy2016rethinking}
\bibinfo{author}{Szegedy, C.}, \bibinfo{author}{Vanhoucke, V.},
  \bibinfo{author}{Ioffe, S.}, \bibinfo{author}{Shlens, J.} \&
  \bibinfo{author}{Wojna, Z.}
\newblock \bibinfo{title}{Rethinking the inception architecture for computer
  vision}.
\newblock In \emph{\bibinfo{booktitle}{Proceedings of the IEEE conference on
  computer vision and pattern recognition}}, \bibinfo{pages}{2818--2826}
  (\bibinfo{year}{2016}).

\bibitem{xie2017aggregated}
\bibinfo{author}{Xie, S.}, \bibinfo{author}{Girshick, R.},
  \bibinfo{author}{Doll{\'a}r, P.}, \bibinfo{author}{Tu, Z.} \&
  \bibinfo{author}{He, K.}
\newblock \bibinfo{title}{Aggregated residual transformations for deep neural
  networks}.
\newblock In \emph{\bibinfo{booktitle}{Proceedings of the IEEE conference on
  computer vision and pattern recognition}}, \bibinfo{pages}{1492--1500}
  (\bibinfo{year}{2017}).

\bibitem{dosovitskiy2020image}
\bibinfo{author}{Dosovitskiy, A.} \emph{et~al.}
\newblock \bibinfo{journal}{\bibinfo{title}{An image is worth 16x16 words:
  Transformers for image recognition at scale}}.
\newblock {\emph{\JournalTitle{arXiv preprint arXiv:2010.11929}}}
  (\bibinfo{year}{2020}).

\bibitem{huang2017densely}
\bibinfo{author}{Huang, G.}, \bibinfo{author}{Liu, Z.}, \bibinfo{author}{Van
  Der~Maaten, L.} \& \bibinfo{author}{Weinberger, K.~Q.}
\newblock \bibinfo{title}{Densely connected convolutional networks}.
\newblock In \emph{\bibinfo{booktitle}{Proceedings of the IEEE conference on
  computer vision and pattern recognition}}, \bibinfo{pages}{4700--4708}
  (\bibinfo{year}{2017}).

\bibitem{lin2017focal}
\bibinfo{author}{Lin, T.-Y.}, \bibinfo{author}{Goyal, P.},
  \bibinfo{author}{Girshick, R.}, \bibinfo{author}{He, K.} \&
  \bibinfo{author}{Doll{\'a}r, P.}
\newblock \bibinfo{title}{Focal loss for dense object detection}.
\newblock In \emph{\bibinfo{booktitle}{Proceedings of the IEEE international
  conference on computer vision}}, \bibinfo{pages}{2980--2988}
  (\bibinfo{year}{2017}).

\bibitem{loshchilov2017decoupled}
\bibinfo{author}{Loshchilov, I.} \& \bibinfo{author}{Hutter, F.}
\newblock \bibinfo{journal}{\bibinfo{title}{Decoupled weight decay
  regularization}}.
\newblock {\emph{\JournalTitle{arXiv preprint arXiv:1711.05101}}}
  (\bibinfo{year}{2017}).

\bibitem{kingma2014adam}
\bibinfo{author}{Kingma, D.~P.} \& \bibinfo{author}{Ba, J.}
\newblock \bibinfo{journal}{\bibinfo{title}{Adam: A method for stochastic
  optimization}}.
\newblock {\emph{\JournalTitle{arXiv preprint arXiv:1412.6980}}}
  (\bibinfo{year}{2014}).

\bibitem{wiesenfarth2021methods}
\bibinfo{author}{Wiesenfarth, M.} \emph{et~al.}
\newblock \bibinfo{journal}{\bibinfo{title}{Methods and open-source toolkit for
  analyzing and visualizing challenge results}}.
\newblock {\emph{\JournalTitle{Scientific reports}}}
  \textbf{\bibinfo{volume}{11}}, \bibinfo{pages}{1--15} (\bibinfo{year}{2021}).

\bibitem{ganaie2021ensemble}
\bibinfo{author}{Ganaie, M.~A.}, \bibinfo{author}{Hu, M.} \emph{et~al.}
\newblock \bibinfo{journal}{\bibinfo{title}{Ensemble deep learning: A review}}.
\newblock {\emph{\JournalTitle{arXiv preprint arXiv:2104.02395}}}
  (\bibinfo{year}{2021}).

\bibitem{xie2016melanoma}
\bibinfo{author}{Xie, F.} \emph{et~al.}
\newblock \bibinfo{journal}{\bibinfo{title}{Melanoma classification on
  dermoscopy images using a neural network ensemble model}}.
\newblock {\emph{\JournalTitle{IEEE transactions on medical imaging}}}
  \textbf{\bibinfo{volume}{36}}, \bibinfo{pages}{849--858}
  (\bibinfo{year}{2016}).

\bibitem{khened2019fully}
\bibinfo{author}{Khened, M.}, \bibinfo{author}{Kollerathu, V.~A.} \&
  \bibinfo{author}{Krishnamurthi, G.}
\newblock \bibinfo{journal}{\bibinfo{title}{Fully convolutional multi-scale
  residual densenets for cardiac segmentation and automated cardiac diagnosis
  using ensemble of classifiers}}.
\newblock {\emph{\JournalTitle{Medical image analysis}}}
  \textbf{\bibinfo{volume}{51}}, \bibinfo{pages}{21--45}
  (\bibinfo{year}{2019}).

\bibitem{ouali2020semi}
\bibinfo{author}{Ouali, Y.}, \bibinfo{author}{Hudelot, C.} \&
  \bibinfo{author}{Tami, M.}
\newblock \bibinfo{title}{Semi-supervised semantic segmentation with
  cross-consistency training}.
\newblock In \emph{\bibinfo{booktitle}{Proceedings of the IEEE/CVF Conference
  on Computer Vision and Pattern Recognition}}, \bibinfo{pages}{12674--12684}
  (\bibinfo{year}{2020}).

\bibitem{bortsova2019semi}
\bibinfo{author}{Bortsova, G.}, \bibinfo{author}{Dubost, F.},
  \bibinfo{author}{Hogeweg, L.}, \bibinfo{author}{Katramados, I.} \&
  \bibinfo{author}{Bruijne, M.~d.}
\newblock \bibinfo{title}{Semi-supervised medical image segmentation via
  learning consistency under transformations}.
\newblock In \emph{\bibinfo{booktitle}{International Conference on Medical
  Image Computing and Computer-Assisted Intervention}},
  \bibinfo{pages}{810--818} (\bibinfo{organization}{Springer},
  \bibinfo{year}{2019}).

\bibitem{bertels2019optimizing}
\bibinfo{author}{Bertels, J.} \emph{et~al.}
\newblock \bibinfo{title}{Optimizing the dice score and jaccard index for
  medical image segmentation: Theory and practice}.
\newblock In \emph{\bibinfo{booktitle}{International conference on medical
  image computing and computer-assisted intervention}},
  \bibinfo{pages}{92--100} (\bibinfo{organization}{Springer},
  \bibinfo{year}{2019}).

\bibitem{khalid2021widefield}
\bibinfo{author}{Khalid, H.} \emph{et~al.}
\newblock \bibinfo{journal}{\bibinfo{title}{Widefield optical coherence
  tomography angiography for early detection and objective evaluation of
  proliferative diabetic retinopathy}}.
\newblock {\emph{\JournalTitle{British Journal of Ophthalmology}}}
  \textbf{\bibinfo{volume}{105}}, \bibinfo{pages}{118--123}
  (\bibinfo{year}{2021}).

\bibitem{cui2021comparison}
\bibinfo{author}{Cui, Y.} \emph{et~al.}
\newblock \bibinfo{journal}{\bibinfo{title}{Comparison of widefield
  swept-source optical coherence tomography angiography with ultra-widefield
  colour fundus photography and fluorescein angiography for detection of
  lesions in diabetic retinopathy}}.
\newblock {\emph{\JournalTitle{British Journal of Ophthalmology}}}
  \textbf{\bibinfo{volume}{105}}, \bibinfo{pages}{577--581}
  (\bibinfo{year}{2021}).

\bibitem{pichi2020wide}
\bibinfo{author}{Pichi, F.} \emph{et~al.}
\newblock \bibinfo{journal}{\bibinfo{title}{Wide-field optical coherence
  tomography angiography for the detection of proliferative diabetic
  retinopathy}}.
\newblock {\emph{\JournalTitle{Graefe's Archive for Clinical and Experimental
  Ophthalmology}}} \textbf{\bibinfo{volume}{258}}, \bibinfo{pages}{1901--1909}
  (\bibinfo{year}{2020}).

\end{thebibliography}
\end{document}